\documentclass[preprint,aps]{revtex4}
\usepackage{epsfig}
\begin{document}
\title{Theory of Analogous Force on Number Sets}
\author{Enrique Canessa\footnote{E-mail: canessae@ictp.trieste.it}}
\affiliation{The Abdus Salam International Centre for Theoretical Physics,
Trieste, Italy
\vspace{2cm}
}

\begin{abstract}
A general statistical thermodynamic theory that considers
given sequences of $x$-integers to play the role of particles 
of known type in an isolated elastic system is proposed.  
By also considering some explicit discrete probability distributions 
$p_{x}$ for natural numbers, we claim that they lead to a better
understanding of probabilistic laws associated with number theory.
Sequences of numbers are treated as the size measure of finite sets.
By considering $p_{x}$ to describe complex phenomena, the
theory leads to derive a distinct analogous force $f_{x}$ on number
sets proportional to 
$\left( \frac{\partial p_{x}}{\partial x} \right)_{T}$
at an analogous system temperature $T$.  In particular, this yields 
to an understanding of the uneven distribution of integers of random
sets in terms of analogous scale invariance and a screened inverse
square force acting on the significant digits.  
The theory also allows to establish recursion relations to predict
sequences of Fibonacci numbers and to give an answer to the
interesting theoretical question of the appearance of the 
Benford's law in Fibonacci numbers.  A possible relevance
to prime numbers is also analyzed.

\vspace{1.0cm}
\end{abstract}
\maketitle

\section{Introduction}
There are suggestive parallels between various aspects of 
number theory and physics phenomena. The connection often
involves affinity with functions arising in statistical
mechanics \cite{Ste92}. For example, a connection between 
given number sets and the thermodynamic formalism has been
established for the multifractal spectrum via a Legendre
transform \cite{Lee88} and the statistical mechanics 
applications for prime numbers via the Riemann partition
function \cite{Wol99}.  Bose condensation has been indirectly
linked to the zeta function in \cite{Jul90}.  Primes as such
have also attracted recent applications in acoustics and
dynamical systems \cite{Dah01}.  Furthermore, the existence
of an imaginary quantum mechanical potential for the prime
numbers has been idealized in \cite{Mus97} after some energy
considerations.  In a similar context, additive number theory
in terms of scale invariance has been used to study the origin
and form of the uneven distribution of integers of many random
data sets \cite{Hil98,Tol00,Adh68,Math,Got02,Sny02}.  

In this paper we introduce an alternative approach which 
considers sets of $x$-natural numbers to play the role of
particles of known type in an isolated elastic system and
assume that their discrete probability distribution can be
split as the product of some functions.  By applying
thermodynamics techniques, we develop an analogous statistical
mechanics version of the average energy and energy states of
arbitrary number sets. 

From the Helmoltz free energy in a reversible isothermal 
process, a relation for an analogous tensile force $F>0$ 
is derived by treating $x$ as variations in length in the 
direction of $F$ and considering the number system as a 
elastic specimen (body).  In particular, our theory allows
to give an alternative physics answer to the uneven distribution
of integers in terms of a screened inverse square force which
ideally may act on these numbers.  It also allows to establish
second order polynomial recursions to predict sequences of 
integers following generalized Fibonacci numbers.  We
also analyze a possible application of the distinct forces 
to primes. 

Hereafter, the word {\it x-number} is used to refer to 
quantities which are integers and positive.  

\section{Underlying Theory}

\subsection{A priori probability postulate}
Let us define the discrete probability distribution of 
finding a certain positive natural number $x$ as
\begin{equation}\label{eq:probability}
p_{x}=g_{1}(x)\; g_{2}(x) \;\;\; ,
\end{equation}
with $g_{1}(x) >0$ and $0< g_{2}(x)\le 1$ and
satisfying the normalization $\sum_{x}p_{x} = 1$.
Such discrete probability distribution functions 
(and also discrete mixtures of them), used to describe complex 
phenomena, are given by the product of two different functions.  
The real function $g_{2}(x)$ here represents the set of digital
frequencies which remain fixed under base changes given by the
real function $g_{1}(x)$.  

This factorization may be seen somewhat arbitrary, however
it is based on composed functions which include the 
Binomial(A,B), Furry or Geometrical, inversed Logarithmic, Pascal
or Polya and Poisson distributions.  Furthermore, a probability 
decomposition into (two) functions is no new.  This type of factorization 
appears formally, {\it e.g.}, in the analysis of stochastic processes 
on graphs.  According to the well-known Hammersley-Clifford Theorem, a
probability function $p(x)$ is formed by a normalized product of positive
functions on cliques ({\it i.e.}, fully connected subgraph) of the graph
$\cal{G}$ when the stochastic process $x$ is Markov relative to $\cal{G}$.

To simplify notation in all later sums and products the variable $x$ 
represents an upper bound (or cutoff).

\subsection{Thermodynamic Analogy}
Using the above probability postulate the information theoretical
entropy $S_{I}=-\sum_{x} p_{x} \; ln \; p_{x}$ then becomes
\begin{equation}\label{eq:entropy}
S_{I} = - \sum_{x} g_{1}(x)\; g_{2}(x) \; \{ \;  ln \; g_{1}(x)  
         + \; ln \; g_{2}(x) \; \} \;\;\; .
\end{equation}

In order to make a thermodynamic analogy, let us consider next
the Helmholtz free energy $A=E-TS$ assuming an analogous temperature $T$ and
volume $V$ to be independent variables, with $E$ the analogous internal energy
of number sets.  Because of the close relationship of Eq.(\ref{eq:entropy})
to the thermodynamic entropy $S$, we can readily identify
\begin{eqnarray}
A/kT & \rightarrow & \sum_{x} p_{x} \; ln \; g_{1}(x) \;\;\; ,
                      \label{eq:helmholtz} \\
E/kT & \rightarrow & - \sum_{x} p_{x} \; ln \; 
               g_{2}(x) \;\;\; .  \label{eq:energy}
\end{eqnarray}
with $k$ the Boltzmann constant.  Reduced units are used throughout.
The choice of the latter quantities may be seen quite arbitrary,
but as mentioned in the conclusion these can be justified in 
analogy with standard thermodynamics results.

In this case, for the correct application of conventional thermodynamics
the normalization of $p_{x}$ is used to define both the Boltzmann and
the Shannon entropies.

For 'non-interacting' number sets $x>1$, we can get an approximated statistical
mechanics version of the average energy as the sum of individual contributions 
$E \rightarrow \bar{E} = \sum_{x} \bar{n}_{x} \epsilon_{x}$.  The total 
number of positive natural number $x$'s is given by 
$N = \sum_{x} \bar{n}_{x}$ and $\epsilon_{x}$
represents their analogous energy states.  Thus, from Eqs.(\ref{eq:probability})
and (\ref{eq:energy}) we further identify
\begin{eqnarray}
\bar{n}_{x} & \rightarrow & - ln \; g_{2}(x)  \; \ge 0 \;\;\; , \label{eq:nx} \\ 
\epsilon_{x}/kT & \rightarrow & g_{1}(x)\; g_{2}(x) \; = \; p_{x} 
     \; \ge 0 \;\;\; .  \label{eq:energyx}
\end{eqnarray}

The number of states available to a molecular system is given by
the partition function $\Omega$ which satisfies the relation 
$S=k \; ln \; \Omega (N,V,E)$ in the microcanonical ensemble and relates 
the canonical partition function by $Q(N,V,T) =\Omega (N,V,E) e^{-E/kT}$.  
Therefore, from Eqs.(\ref{eq:entropy}), (\ref{eq:energy}) and (\ref{eq:energyx}) 
we also have in analogy
\begin{equation}\label{eq:partition}
Q_{x} = \prod_{x} [g_{1}(x)]^{- p_{x}} = \prod_{x} [g_{1}(x)]^{- \epsilon_{x}/kT} \;\;\; .
\end{equation}
As expected, the form of our canonical partition function depends on the number
probability distribution of Eq.(\ref{eq:probability}).

\subsection{Analogous Force}
To derive a relation for an analogous external tension which we may ideally
apply on number sets in a gedanken experiment, let us treat $x$ as a
variation in length in the direction of a tensile force $f_{x}$ and the
number system as an elastic body.
As a first approximation, the required relation is then
obtainable directly from the first and second laws of thermodynamics in a
reversible isothermal process \cite{Tre75}.  For a change at constant $T$, 
it follows that $dA = dW = f\; d\ell$, where $W$ is the work done on the 
system by its surroundings in a small elongation $d\ell$.  Hence, by partial
differentiation of Eq.(\ref{eq:helmholtz}),
\begin{equation}\label{eq:force}
f_{x}=\left( \frac{\partial A}{\partial x}\right)_{T} = 
  \frac{\epsilon_{x}}{g_{1}(x)} \left( \frac{\partial g_{1}(x)}{\partial x}\right)_{T} +
 ln \; g_{1}(x) \; \left( \frac{\partial \epsilon_{x}}{\partial x}\right)_{T} \;\;\; .
\end{equation}
Since the free energy depends on the upper cutoff $x$ then the derivative
was taken in terms of that variable, {\i i.e.} the derivative of the last 
term in the sums.
If $f_{x}<0$, this corresponds to an analogous compressive force on the number
system.  We shall show next that this new tensile force may indeed be useful 
to characterize number sets.

\section{Probability Distributions and Related Functions}
\subsection{First Digits Patterns: $g_{1}(x)\equiv c_{0}$ , $g_{2}(x)\equiv f [ln \; x]$}
In terms of the first digits of unbiased random samples from random
distributions, the so-called Benford's law follows the logarithmic distribution 
\begin{equation}\label{eq:benford}
p_{x} = log_{10} \; (1+\frac{1}{x}) \equiv c_{0} \;
         ln \; \gamma (x)  \;\;\; ,  
\end{equation}
with $1/c_{0} \equiv ln \; 10$, $\gamma (x)=1+\frac{1}{x}$ and $x=1, 2,\cdots$.  
This empirical law deals with
the distribution of first digits in numbers in which significant positive
digits are not uniformly distributed, but smaller digits and smaller
combinations of significant digits are favored.

The sum of probabilities in this case is 
\begin{equation}\label{eq:beta}
\sum_{x}p_{x} \equiv \beta_{x} = c_{0}\sum_{x} ln \; \gamma (x) = c_{0} \; ln \; \prod_{x} (\frac{x+1}{x}) 
          = c_{0} \; ln \; (1+x) \;\;\; .
\end{equation}
This result implies that $p_{x}$ is not a proper probability since it imposes 
some maximum value \cite{Hil98}.

From Eq.(\ref{eq:partition}) the evaluation of the analogous partition
function is straight forward, namely
\begin{equation}
Q_{x} = \prod_{x} \gamma (x)^{- c_{0} \; ln \; c_{0}} 
      = (1+x)^{- c_{0} \; ln \; c_{0}} \;\;\; .
\end{equation}

According to Eq.(\ref{eq:force}), our analytical expression for the 
analogous force acting on this number set becomes 
\begin{eqnarray}\label{eq:force_approx}
 \frac{f_{x}}{c_{0} \; ln \; c_{0}} &  =  &  - \; \frac{1}{x+x^{2}} \;\;\; .
\end{eqnarray}
It is from this base-invariant behavior of a distinct $f_{x}$ that we can
make an attempt to understand the significant digit distribution of 
Eq.(\ref{eq:benford}) in terms of an screened inverse square force.

\subsection{Relevance to Primes: $g_{1}(x)\equiv e^{-1}$ , $g_{2}(x)\equiv x^{-\sigma}$}
If we generalize the above to a power law behavior for the sequence of
primes only greater than one with an exponent $\sigma >1$ ,
{\it i.e.}, $p_{x_{p}} = 1/(e\; x_{p}^{\sigma})$,
then the class of analogous forces on prime numbers becomes
\begin{equation}
f_{x_{p}} = \frac{\sigma}{x_{p}^{\sigma+1}} \;\;\; ,
\end{equation}
-a fingerprint for primes.

Using Eq.(\ref{eq:partition}), the analogous partition function approaches
\begin{equation}
\frac{Q_{x_{p}}}{e^{1/e}} \approx \prod_{x_{p}>1} \frac{1}{1-x_{p}^{-\sigma}}
         = \sum_{x} \frac{1}{x^{\sigma}} \;\;\; .
\end{equation}

Another possible relation of our statistical thermodynamic approach to prime
numbers follows by considering the average system energy.  In the limit
of Eq.(\ref{eq:force_approx}), this takes the form
\begin{equation}
\frac{E}{c_{0}} \approx \sum_{x} \frac{ln \; x}{x} \;\;\; .
\end{equation}
In principle this could be asymptotic to $\sum_{x} \frac{1}{\pi (x)}$,
where the counting function $\pi (x)$ denotes the number of prime numbers
$x_{p}$ that do not exceed $x$ \cite{Ten00}.

\section{Discussion}

\subsection{Benford's law, Analogous Force and Thermodynamics}
\begin{figure}
\epsfxsize=12cm
\epsfbox{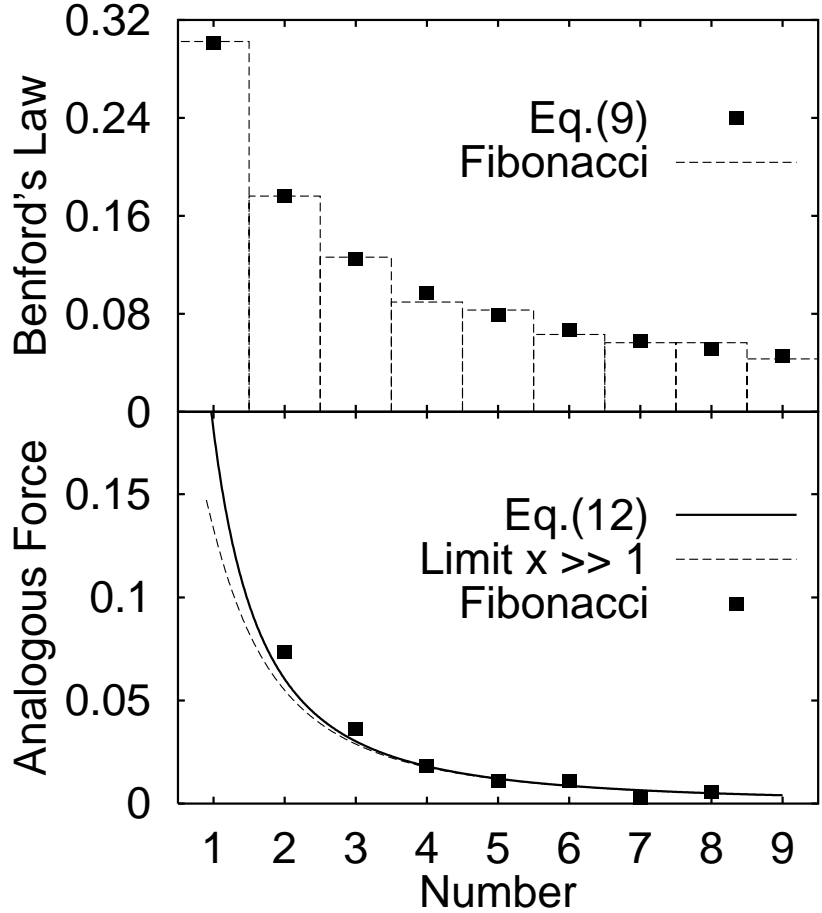}
\caption{Probability distribution of, and distinct force on, Fibonacci
numbers.}
\label{Fig.1}
\end{figure}

As an example, in Fig.\ref{Fig.1} we plot the probability distribution and
our distinct force on (up to 300) numbers from Fibonacci series
and compare results with the present analogous statistical
thermodynamics approach.  The theoretical calculation of the analogous
force is obtained using Eq.(\ref{eq:force_approx}) (and its large-$x$ 
behavior).
The numerical calculation of $f_{x}$ for Fibonacci numbers is also rather 
simple.  This is derived using Eq.(\ref{eq:force}) and from the fact that
$c_{x}$ is constant.  By the two-point central formula we approximate
\begin{equation}\label{eq:force_approx2}
f_{x}/ln \; c_{0} = \left( \frac{\partial p_{x}}{\partial x}\right)_{T}
\approx \frac{1}{2}(p_{x+1} - p_{x-1}) \;\;\; .
\end{equation}
This means that the probability distribution $p_{x}$ is related to an
analogous local potential field at each number ({\it i.e.}, particle)
as in the well known diffusion limited aggregation model \cite{Wit81}.
Assuming a sort of superposition principle, the total analogous force 
$F\equiv\sum_{x} f_{x}$ can also be readily evaluated.  From 
Eq.(\ref{eq:force_approx}) we obtain $F/c_{0} =- ln \; c_{0} / \gamma (x)$. 

Since $c_{0} \approx 0.4343$ in logarithm base $10$, we then
have an analogous system entropy $S>0$, the analogous Helmholtz free energy
$A<0$, the analogous partition function $Q_{x}>0$, an analogous {\it expansive}
force on the integers $f_{x}>0$ and the sum of probabilities $\beta_{x}>0$.
Interestingly, from Eq.(\ref{eq:beta}) we also have that this sum only becomes
normalized at $x=9$ ({\it i.e.}, a decadelong). 
On the other hand, the increase in the analogous
entropy as $x$ increases follows the natural tendency of thermodynamic
systems to move toward disordered states having more microscopic states
characterized by our analogous macroscopic variables.

Besides, the particular $\gamma (x)$ under consideration allows to derive
a measure for an analogous inverse temperature of the integers system.  From
Eqs.(\ref{eq:entropy}), (\ref{eq:energy}) and (\ref{eq:benford}), we thus
approximate
\begin{equation}\label{eq:temperature}
\frac{\Delta (S/k)}{\Delta (E/kT)} = 1 +
     \frac{ln \; c_{0}}{ln \; (ln \; \gamma (x) )} \;\;\; ,
\end{equation}
where $\Delta$ represents functions difference between consecutive number
fluctuations. 

For $x>>1$, Eq.(\ref{eq:force_approx}) implies that
$p_{x} \approx c_{0}x^{-1}$.  Such a power
law behavior with exponent one is the solution of the functional form
$p_{x'}=p_{ax}=A_{a}p_{x}$ which may correspond to an analogous
scale invariance \cite{Math}.

The results shown in the figure allows to argue that our simple
physics-based approach is not merely speculative.  Indeed it yields to a
 different and more complete understanding of Benford's law
than that of previous attempts \cite{Adh68,Math,Got02,Sny02}.  It includes
scale invariance, frequencies behavior under base changes and explanation
for a logarithmic law distribution besides the analogous thermodynamic
characterization of given number sets such as time series or primes.

\subsection{Screening}
For $x >> 1$, the force $\frac{f_{x}}{c_{0} \; ln \; c_{0}}$ can be
approximated as $- \; \frac{e^{- \frac{1}{x}}}{x^{2}}$, so we could
consider the exponential term as being screening.  The presence of 
$exp(-1/x)$ screening in this analogous inverse-squared force
implies that the second leading digits ({\it i.e.}, $ 10 \le x \le 99$)
are much more uniformly distributed than the first digits, and the third than
the second, and so on according to observations \cite{Hil98}.  This is
reflected by the numerical differences $f_{1}-f_{9}=0.17708$,
$f_{10}-f_{99}=0.00326$, $f_{100}-f_{999}=0.00004$.
The main message is that the asymmetry in favor of smaller significant digits
or combinations of significant digits can be understood in terms of analogous
screened inverse square force acting on the number sets.

Indeed screening could be a way to unveil the fact that significant digits
are {\it not independent}, but that instead knowledge of one digit affects
the likelihood of another.  Furthermore, since screening is
a consequence of having the base invariant probability distribution of
Eq.(\ref{eq:probability}), then the choice of digital frequencies bounded
by $1 < 1 + \frac{1}{x} \le e$ leads to support a logarithmic law
distribution.

\subsection{Recursions: A Possible Scenario for Number Prediction}

From a physical point of view, the equilibrium of forces within 
defect-free plate under external shear can be approximated by
the discrete Laplace equation\cite{Tak85,Can91,Can92}
\begin{equation}\label{eq:gkuk}
\sum_{k=1}^{4} \; G_{k}\Delta u_{k}  = 0 \;\;\;
\end{equation}
where $\Delta u_{k} = u_{k}(i) - u(i)$ denotes relative displacements
of the $i$th lattice point with respect to its nearest neighbors
and $G_{k}$ is the corresponding modulus of rigidity of
the bond connecting the $i$th lattice point to its neighbor.

From Eq.(\ref{eq:gkuk}) it follows then that for a 1D system
\begin{equation}
u(i) = \hat{G}_{1} u_{1}(i) + \hat{G}_{2} u_{2}(i) \;\;\; ,
\end{equation}
where $\hat{G}_{i} = G_{i}/\sum_{i}G_{i}\;\; (i=1,2)$.

It is not unreasonable to assume that the moduli of rigidity
of our analogous elastic system are proportional 
to each other, at least prior to any equivalent breakdown
condition which may prevail.  Therefore if we consider
$u_{1}(i) \propto (G_{2}/G_{1}) u_{2}(i-1)$, then we have that
all displacements $u_{2}$, $u_{1}$ and $u$ follow
generalized fibonacci sequences of the form \cite{Gol77}
\begin{equation}
{\cal U}(i+2) = {\cal U}(i) + {\cal U}(i+1) \;\;\; .
\end{equation}

Within our thermodynamics framework, the sequences of $x$-integers 
played the role of analogous 'non-interacting' particles.  Besides, 
their magnitude may be seen as representing the above effective 
(increasing) displacements, say expected values 
$x = {\cal E}[{\cal U}]$, at different times $i$.  
The analogous force may be in turn seen as
an applied shear acting on our 1D isolated elastic 
system of finite size $x$, which is being pulled out (or in).

Therefore following these ideas from fracture phenomena, a possible scenario
to deduce second order polynomial recursions leading to predict
sequences of numbers becomes possible. 

\section{Concluding Remarks}
In this paper we have introduced a quite abstract statistical
mechanical system whose states are represented by natural numbers.
By considering some explicit probability distributions for the
natural numbers, we claim that they lead to a better understanding
of probabilistic laws associated with number theory.

Within this kind of thermodynamic formalism we have investigated
probability distributions which lead to Benford's law.
Sequences of numbers have been treated as the size measure of finite sets
and, in analogy with statistical thermodynamics, they have been assumed
to play the role of particles of known type in an isolated elastic system.

The relation to physics is rather in the thermodynamical formalism
and terminology for number theoretical problems.
The assignment of statistical quantities here proposed are not arbitrary.
Our analogies for the considered functions, {\it e.g.}, Helmoltz free
energy, {\it etc}, follows the similar thermodynamical approach as
proposed within Multifractal physics.
The expression of the Shannon entropy in Eq.(\ref{eq:entropy}) split 
into two terms with the identification of one of the terms
with an analogous free energy and the other with an analogous internal
energy can be justified (although not rigorously) in terms of
the form of our canonical partition function which depends on the number
probability distribution of Eq.(\ref{eq:probability}) as may be expected.
Furthermore as derived from such identification, we can argue that the relation in
Eq.(\ref{eq:temperature}) tends to unity in the large $x$-limit in 
analogy with the standard thermodynamic definition 
$T=\left( \frac{\partial E}{\partial S}\right)_{V,N}$.

Though this scheme may be seen as rather abstract, we have shown that it leads
to very interesting connections with statistical thermodynamics and allows
to interpret physically the observed probability distribution of first
digits in terms of a stretching force (tension) which ideally may pull
integers number sets along analogous lengths.

The introduction of an analogous force within our formalism has been
done on a formal level.  The new insights gained by such a formal introduction
and its utility include, in particular, the study of the dynamics of given
number sets by establishing second order polynomial recursions to predict
sequences of generalized Fibonacci numbers. This gives an answer to the
interesting theoretical question of the appearance of the Benford's law
in Fibonacci numbers.

Following recent interest in Benford's law \cite{Hil98}, we believe the new
force may find application to detect fraud in financial data \cite{Nig96},
by looking for anomalies in their analogous macroscopic thermodynamic variables, 
in the use of logarithmic computers to speed calculations \cite{Knu81}, by
analyzing the screening behavior of the analogous inverse square forces
on floating points, or in the characterization of the hopcount in Internet
\cite{Van00}. 

\section*{Acknowledgement}
The author would like to thank the ARPL Group of the Abdus Salam ICTP, Trieste,
for hospitality and support.

\end{document}